\documentclass[a4paper]{jpconf}
\usepackage{graphicx}
\usepackage{subfigure}
\usepackage{amsmath}

\def\endthebibliography{%
	\def\@noitemerr{\@latex@warning{Empty `thebibliography' environment}}%
	\endlist
}

\begin{document}
\title{Spontaneous atomic crystallization via diffractive dephasing in optical cavities}

\author{A. Costa Boquete, G. Baio, G.R.M. Robb, G.-L. Oppo, P. Griffin, and T. Ackemann}

\address{SUPA and Department of Physics, University of Strathclyde, Glasgow G4 0NG, Scotland, UK}

\ead{adrian.costa-boquete@strath.ac.uk}

\begin{abstract}
The design of an experiment on the spontaneous crystallization of a laser-cooled, but thermal atomic cloud into a hexagonally structured phase is discussed. Atomic interaction is mediated by the dipole potential of an optical lattice formed spontaneously in a multi-mode degenerate cavity from single-mode longitudinal pumping. The length scale of the structure is given by the diffractive dephasing between the spontaneous sidebands and the on-axis pump. A linear stability indicates that the transition can be observed in a cavity of moderate finesse compatible with having the cavity mirrors outside the vacuum cell. A new anti-reflected cell has been assembled for this purpose.
\end{abstract}

\section{Introduction}
Recent years have witnessed considerable interest in achieving effective interaction between internal or external degrees of freedom of atoms via light modes to explore new self-organized phases and/or the dynamics of complex many-body systems by well-controlled cold atom model systems. A workhorse has been the transversely pumped cavity \cite{ritsch13}, originally proposed in \cite{domokos02}. Self-organization of atoms into a checkerboard lattice has been demonstrated experimentally for laser-cooled thermal \cite{black03} and quantum degenerate samples \cite{baumann10}. In these schemes the spontaneously generated intra-cavity field interferes with the pump field providing a modulated optical potential in which atoms bunch into an optical lattice. The scattering of the pump field by this lattice in turn sustains the intra-cavity field. If the cavity supports only a single spatial mode, the intra-cavity field provides a global coupling between atoms. In order to enable more local coupling there has been considerable interest to explore multi-mode configurations either via degenerate cavities \cite{gopalakrishnan09,kollar17} or single-mode cavities with crossed axes \cite{leonard17}. All these schemes have a limitation in the sense that there are two (or more) distinguished axes which imposes restrictions on the structures which can emerge from self-organization.

In contrast, a longitudinally pumped multi-mode cavity possesses only one distinguished axis and allows for spontaneous breaking of the rotational and translational symmetry in the plane transverse to the pump axis, at least for the idealized case of pumping with a plane wave. This effort builds on intense investigations on optical pattern formation in longitudinally pumped cavities in the 1980-2000s, in which however the emphasis has been on the coupling of optical waves via the nonlinear medium \cite{lugiato87,firth92,tlidi93} and not the coupling of degrees of freedom in atomic matter via the optical fields. Due to some experimental difficulties, experimental realizations using atomic media were confined to few mode situations \cite{lippi93}, whereas  multi-mode situations were explored using semiconductors or liquid crystals \cite{kreuzer90,ackemann00}. In Ref.~\cite{tesio12,tesio14t} a theory was developped to describe optomechanical self-organization of atoms in a longitudinally pumped plano-planar cavity. In these Proceedings, we will briefly review the mechanism and the model, then proceed to a more detailed linear stability analysis indicating that the observation is well within reach of current cold atom technology and then report on the experimental design and the progress on a new vacuum cell for this experiment.

\section{Theoretical analysis}
\subsection{Mechanism of diffractive coupling}\label{sec:mech}
\begin{figure}[b]
\begin{center}
\subfigure[Below threshold]{\includegraphics[width=13cm]{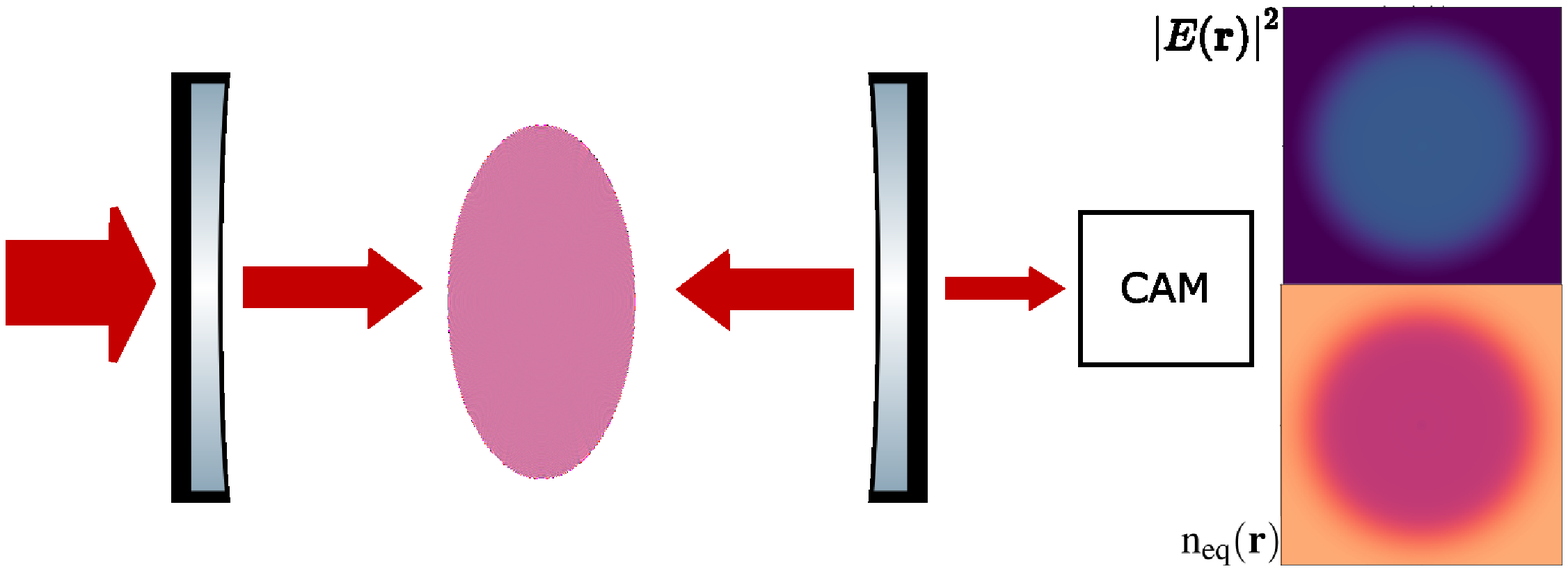}\label{Fig:scheme_a}}
\subfigure[Above threshold]{\includegraphics[width=13cm]{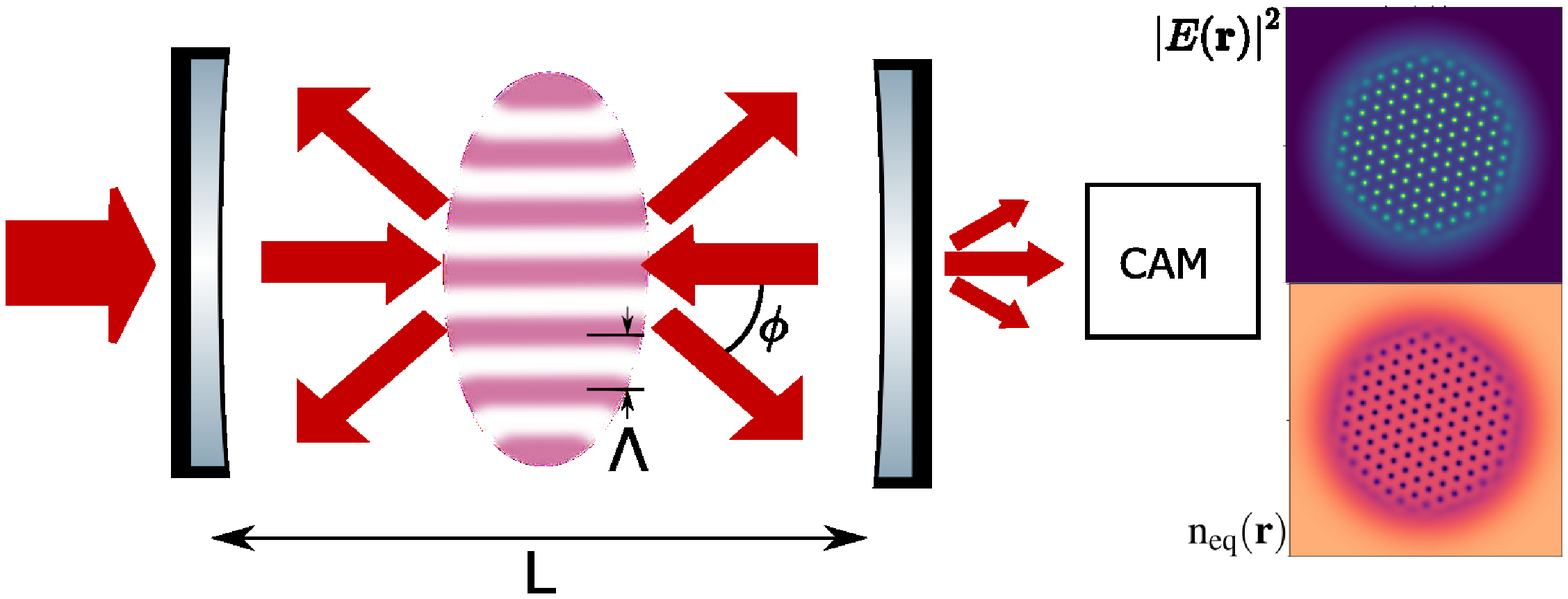}\label{Fig:scheme_b}}
\caption{\label{fig:scheme}Mechanism of self-ordering.
a) A plano-planar cavity with a cloud of atoms in the center is pumped longitudinally, with a beam detuned from the cavity resonance.
The intensity distribution within the cloud is imaged onto a camera monitoring the transmitted field. Below threshold intra-cavity field and atomic density remain homogeneous. The insets show numerical simulations for a super-Gaussian input beam.
b) Above threshold, spontaneously generated spatial sidebands at an angle $\phi$ to the optical axis interfere to drive the atoms into a real-space structure with period $\Lambda$. Pump photons scattered by this strucure sustain the sidebands. This feedback process drives the formation of a self-induced period structure in the atomic density.}
\end{center}
\end{figure}

Fig.~\ref{fig:scheme} shows the scheme analyzed. A plano-planar cavity is pumped longitudinally by an external coherent beam. It is detuned far enough from the atomic transition to avoid absorption and to make the optomechanical nonlinearities the dominant one. Below threshold (Fig.~\ref{Fig:scheme_a}) the pump beam and the atomic cloud are unstructured. If a finite pump beam (a Gaussian in the experiment, a super-Gaussian in the simulations presented) is used, there is some effect on the atomic density distribution. More complex phase structured pumps can also be used to induce rotational effects and atomic transport \cite{baio20a}. In the situation of positive atomic detuning assumed here, atoms are pushed out of the pump beam via the dipole force. It is important to note now that if the pump beam is not only detuned to the atomic transition but also to the cavity, a spatial sideband at the frequency of the pump can be resonant to the cavity as it has a different diffractive phase shift than the on-axis pump. Hence it can experience gain to develop spontaneously. More concretely, this is the case for positive cavity detuning is which the frequency of the pump light is higher than the longitudinal resonance. These sidebands will interfere to form an optical lattice in the atomic cloud (Fig.~\ref{Fig:scheme_b}). This optical lattice will induce bunching of the atoms via dipole forces, i.e.\ the formation of a density modulated structure representing atomic crystallization. Similar to the case of transversely pumped cavities, the positive feedback loop is then closed by scattering of the pump into the sidebands by the atomic structure. For the configuration under study, the simulations below (see also \cite{tesio12,tesio14t}) and analytical considerations \cite{baio20b} predict the formation of hexagonal structures. As atoms are low field seekers for positive atomic detuning, maxima in the intensity correspond to minima in the density and vice versa leading to a hexagonal structure in intensity and a honeycomb structure in the atomic density. Rotational symmetry is spontaneously broken. The finite pump beam confines somewhat the translational symmetry breaking, but in broad enough pump beams also some shot-to-shot variations of peak positions were observed in related systems \cite{labeyrie14}. It should be noted that diffractive dephasing of sidebands relative to a single pump axis can be also utilized for multi-mode self-organization in single-mirror feedback systems \cite{dalessandro91,ackemann01} and counter-propagating beams \cite{grynberg88,firth90,muradyan05}. Both were recently demonstrated to work in cold atoms \cite{greenberg11,labeyrie14,schmittberger16,kresic18,labeyrie18}. However, the transition to a cavity schemes has a huge potential gain in effective interaction strength. It should be also noted that similar feedback loops can be constructured for internal degrees of freedom of the atoms, e.g.\ Zeeman sublevels leading to spontaneous magnetic ordering as demonstrated for single-mirror feedback systems \cite{kresic18,labeyrie18} and meeting recent interest in spin textures in quantum degenerate gases \cite{mivehvar17,jaeger17,landini18}.

The angle $\phi$ of the sidebands with the highest gain and thus the lowest threshold can be calculated from the differential diffractive phase shift they have compared to the pump. The highest gain corresponds to sidebands resonant to the cavity, i.e.\ sidebands whose wavevector projection along the optical axis is identical to the resonant wavenumber of the cavity.
The diffractive phase shift $\delta\varphi(\phi)$ experienced by the sidebands is then given by
\begin{equation}\label{Eq:Projection}
			\delta\varphi(\phi) = k_0 \,\cos{(\phi})\, 2L \approx  k_0\,\left(1-\frac{\phi^2}{2}\right)\, 2L = k_c 2L,
\end{equation}
where $k_0$ is wavenumber of the scattered sideband (and of the pump!), $k_c$ the resonant cavity wavenumber and $L$ the cavity length.

The phase mismatch $\delta\varphi$ between pump and cavity induced by the cavity detuning is expressed in terms of the wavenumbers by

\begin{equation}\label{Eq:phaseshift}
	\delta\varphi = (k_0-k_c)2L.
\end{equation}
Inserting Eq.~(\ref{Eq:Projection}) into (\ref{Eq:phaseshift}) one obtains a direct relation between the cavity detuning and the scattering angle:
\begin{equation}\label{Eq:k0}
		\delta\varphi = k_0 \phi^2 L
\end{equation}
It's convenient to introduce the transversal wavevector $k_\perp = 2\pi/\Lambda$, with $ \Lambda $ being the spatial periodicity of the transversal pattern. With $k_\perp = k_0\sin{\phi} \approx k_0 \phi$, one obtains
\begin{equation}
	\delta\varphi = \frac{k_\perp^2}{k_0}L = \dfrac{2\pi L\lambda}{\Lambda^2}
\end{equation}
or
\begin{equation}
	\Lambda = \sqrt{\frac{2\pi L\lambda}{\delta\varphi}},
\end{equation}
where the expression for the transverse pattern periodicity has been made explicit. It only depends on the phase mismatch, the cavity length and the beam wavelength. As dicussed below in Sec.~\ref{sec:cavity}, the minimum cavity length dictated by the vacuum apparatus will be on the order of $L\approx 75-100$~cm. For a pump with $\lambda\approx 780$ nm, this gives a numerator of the order of 700 $\mu$m, i.e.\ rather large length scales for cavity detunings smaller than 1. This and diffraction losses indicate that a realization of the theoretical proposal via a straightforward plano-planar cavity is probably not feasible. Instead a planar-cavity with intra-cavity lenses or mode-degenerate cavities like the confocal or hemispherical cavity can be used to obtain an effective diffraction length smaller than the physical cavity length. For a close to confocal cavity it can be shown that the deviation from confocality $l$ represents an effective diffraction length \cite{staliunas97c}. Taking for example $l\approx 100$~$\mu$m yields a prefactor of about $20$~$\mu$m, which makes pattern periods of tens to 100 or 200 $\mu$m feasible.

\subsection{The model}
We look at optomechanical effects in an ensemble of Doppler cooled Rb atoms within a single-mode cavity longitudinally pumped by a coherent plane wave source. Neglecting longitudinal effects, the transversal dynamics of the slowly varying cavity envelope $E(\mathbf{r},t)$ is well described by the following paraxial equation in the mean field limit \cite{lugiato88b}:

\begin{equation}
\frac{\partial E(\mathbf{r},t)}{\partial t}=-\kappa(1+i\Theta)E(\mathbf{r},t) + A_{I} - \kappa\frac{C(1+i\Delta)}{1+s(\mathbf{r},t)}\, n(\mathbf{r},t)\, E(\mathbf{r},t) + i\kappa a\nabla_{\mathbf{r}}^2 E(\mathbf{r},t)
\label{eq:field}
\end{equation}

where $\mathbf{r}$ is the transversal coordinate, $\kappa = c {\cal T}/2L$ the cavity linewidth (with $c$ being the speed of light in vacuum and ${\cal T}=\ln{R_eff}$ the effective cavity losses), $A_I$ the pump rate, $\Delta$ the light-atom detuning normalized to half the atomic transition linewidth $\Gamma$ of the Rb atoms, $C = b_0 /[2{\cal T} (1+\Delta^2)]$ the so-called \textit{cooperativity parameter} of the cavity (depending on the optical density at resonance $b_0$), and $\sqrt{a} = \sqrt{l/k_0 {\cal T}}$ the \textit{diffraction length} ($l$ is the effective diffractive length of the cavity discussed in Sec.~\ref{sec:mech}). Note that Eq.~(\ref{eq:field}) includes in principle the nonlinear effect of atomic saturation $s(\mathbf{r},t) = |E(\mathbf{r},t)|^2$ and couples to the atomic density distribution $n(\mathbf{r},t)$. A set of externally controlled optical molasses beams provides velocity damping to the atom center-of-mass dynamics. Therefore, in the limit in which the frictional force overcomes inertia, the dynamics of $n(\mathbf{r},t)$ is governed by the Smoluchowski diffusive equation \cite{saffman08,tesio12}:

\begin{equation}
\frac{\partial n (\mathbf{r},t)}{\partial t}= \beta D\nabla_{\mathbf{r}}\cdot\left[n(\mathbf{r},t)\,\mathbf{f}_{\textrm{dip}}(\mathbf{r},t)\right]+D\nabla_{\mathbf{r}}^{2}n(\mathbf{r},t) ,
\label{eq:smolu}
\end{equation}

where $\beta=(k_{B}T)^{-1}$ and $D$ denotes the diffusivity of the atomic cloud \cite{hodapp95}. The drift term in Eq.~(\ref{eq:smolu}) is generated by the optomechanical dipole force as follows:

\begin{equation}
\mathbf{f}_{\textrm{dip}}(\mathbf{r},t)  = -\frac{\hbar\Gamma\Delta}{4}\frac{\nabla_{\mathbf{r}}s(\mathbf{r},t)}{1+s(\mathbf{r},t)} ,
\label{force}
\end{equation}

The system described by Eqs.~(\ref{eq:field}) and (\ref{eq:smolu}) displays a modulation instability due to a competition of the two nonlinearities, i.e., the electronic and optomechanical (due to the coupling with $n (\mathbf{r},t)$). This last effect is better described in the low saturation (large detuning) limit, where Eq.~(\ref{eq:smolu}) admits a  stationary state provided by a canonical distribution
\begin{equation}
n_{\textrm{eq}}(\mathbf{r})  =\frac{\exp[-\sigma s(\mathbf{r})]}{\int d^2 \mathbf{r}\exp[-\sigma s(\mathbf{r})]}
\label{eq:gibbs}
\end{equation}

with the constant $\sigma = \hbar\Gamma\Delta/4k_B T$. Numerical results arising from Eq. (10) of the coupled intensity and density distribution in such regime are shown in Fig. \ref{Fig:scheme_a} and \ref{Fig:scheme_b}. In the following section we discuss the dependence of the instability threshold from the model parameters as a result of a linear stability analysis.

\subsection{Results of linear stability analysis} \label{sec:lsa}
\begin{figure}[t]
\begin{center}
\includegraphics[width=12cm]{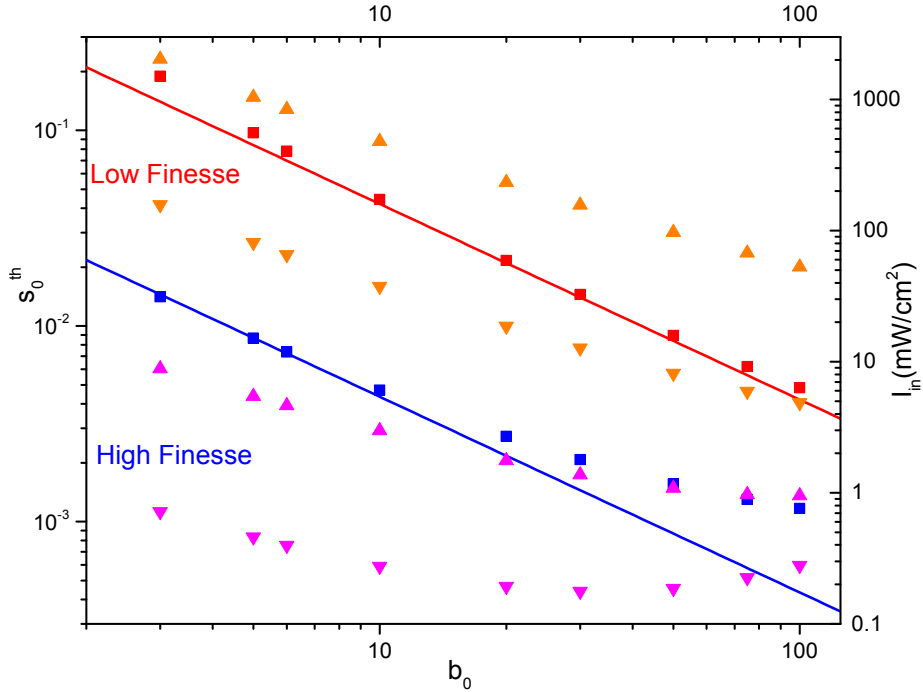}
\caption{\label{fig:lsa}Intra-cavity saturation parameter $s_0^{th}$ vs on-resonance optical density $b_0$ as obtained from a linear stability analysis (squares). The solid line represents the prediction of Eq.~(\ref{eq:optomech}). See text for details. Down triangles represent the extra-cavity pump intensity for an effective detuning of $\Theta_{eff}=-1$ and up triangles the one for $\Theta_{eff}=-5$, calculated for an incoupling mirror with transmission of 0.006. Red and orange data denote low finesse cavity with ${\cal T}=0.1$, blue and magenta high finesse cavity with ${\cal T}=0.01$.}
\end{center}
\end{figure}

The homogeneous steady state distribution for the density in Eq.~(\ref{eq:smolu}) is simply $n_0=1$. Taking the homogeneous intra-cavity intensity $s_0$ as free parameter and analyzing the growth rate of small perturbations of the homogeneous solutions of Eqs.~(\ref{eq:smolu}), (\ref{eq:field}), (\ref{eq:gibbs}) against spatially inhomogeneous perturbations proportional to $\exp{(ik_\perp x)}$, one obtains a $3\times3$-matrix. Threshold correspond to where the determinant of this matrix is zero \cite{tesio14t}. Fig.~\ref{fig:lsa} shows an evaluation of this condition for the parameters ${\cal T} = 0.01$, $\Delta=50$, $T=150$~$\mu$K (or $\sigma=24.3$) and illustrates the dependence of the intra-cavity saturation parameter on the on-resonance optical density $b_0$ as an experimentally easily measurable parameter. For low optical density, it shows a linear dependence in log-log scale with a slope of -1. This is due to the fact that in the strongly detuned regime, the threshold is dominated by the optomechanical nonlinearity, not the electronic 2-level nonlinearity, as the $s_0$ in the denominator of the nonlinear term in Eq.~(\ref{eq:field}) can be neglected. In this limit an analytical expression for the threshold can be found \cite{tesio14t},
\begin{equation} \label{eq:optomech}
s_0^{th}= \frac{1}{C\sigma\Delta}= \frac{1}{b_0}\, \frac{2{\cal T}(1+\Delta^2)}{\sigma\Delta} ,
\end{equation}
which is in very good agreement with the numerical evaluation. For higher optical density, the finesse of the cavity is already affected by absorption which increases the cavity losses.

The threshold saturation parameter does not depend on cavity detuning as $s_0$ is an intra-cavity intensity. For the conversion to the extra-cavity pump intensity, it is assumed here that the cavity detuning is adjusted such that the effective cavity detuning $\Theta_{eff}=\Theta +C\Delta/(1+s_0)$ remains constant at $\Theta_{eff}=-1$ and $\Theta_{eff}=-5$, respectively. From the stationery state of Eq.~(\ref{eq:field}), one can relate $s_0$ to the intensity of the input field $|A_{in}|^2$ (still intra-cavity and scaled) and then via mirror reflectivity and the Rb saturation intensity of $I_{sat}=1.6$~mW/cm$^2$ to an extra-cavity intensity. The results indicate moderate intensities of 1-10~mW/cm$^2$.
    
The spatial period of the structure depends, as explained in Sec.~\ref{sec:mech}, on the cavity detuning. For an effective diffractive length of $l\approx 100$~$\mu$m, one obtains $\Lambda\approx 50\ldots 100$~$\mu$m for the parameter here, which is on the order of the scale demonstrated in single-mirror feedback experiments before. The results indicate that threshold can be reached very comfortably. 

Even with a factor of 10 lower finesse threshold can be reached still with reasonable input intensities of up to 1000 mW/cm$^2$. For a typical input beams with 1~mm radius at the $1/e^2$-point of intensity, a peak intensity of 1000~mW/cm$^2$ corresponds only to 16 mW of input power. For the low finesse cavity it is interesting to note that threshold agrees quite well with the analytical prediction even for high optical density as the additional absorption is low compared to the background losses. However, for low optical densities, there is now a slight deviation from the purely optomechanical limit emerging as a saturation of $s_0\approx 0.2$ is not completely negligible any more.

\section{Experimental considerations}
\subsection{Planned setup and cavity design}\label{sec:cavity}
\begin{figure}[h]
	\begin{minipage}{17pc}
		\includegraphics[width=17pc]{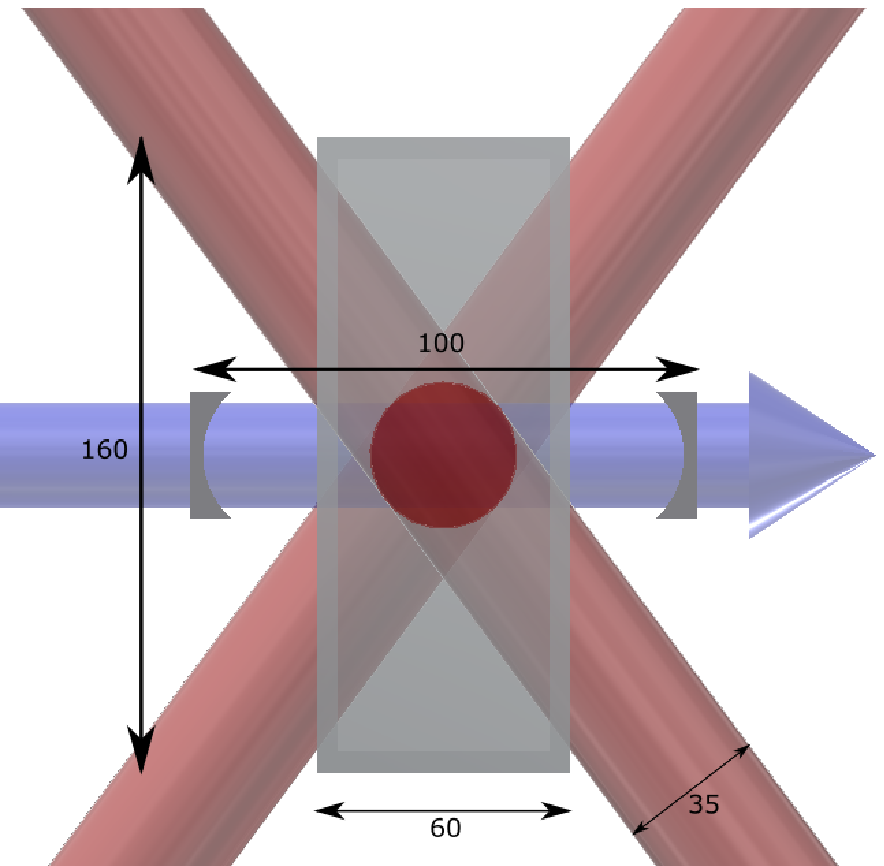}
		\caption{\label{Fig:cell_scheme}Diagram of the setup of the cell, with the two cavity mirrors outside of it, plus the six trapping beams (red) and the cavity pump beam (blue). Distances are measured in mm.} \end{minipage}\hspace{5pc}%
	\begin{minipage}{14pc}
		\includegraphics[width=14pc]{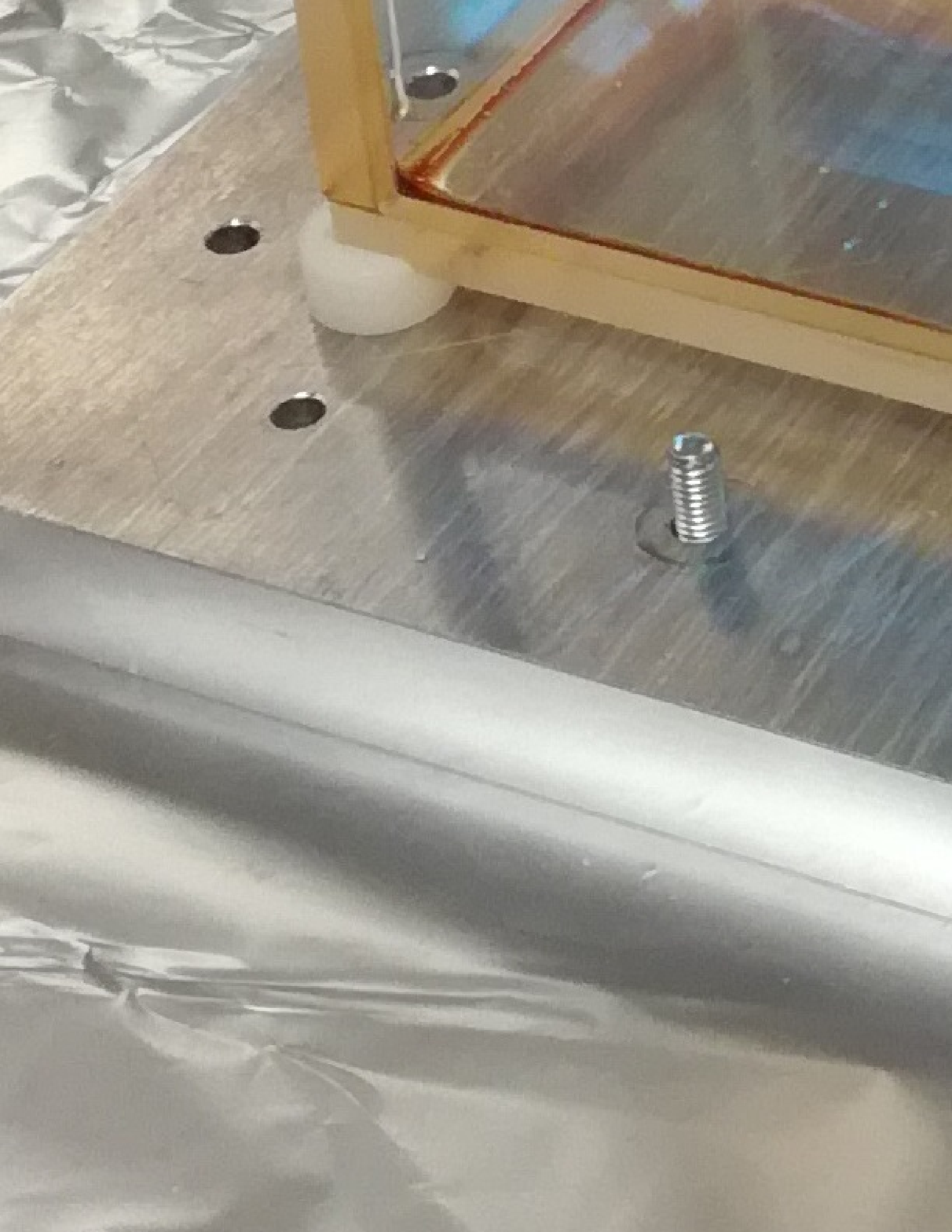}
		\caption{\label{Fig:cell_pic}Picture of the finished cell, before placing in the vacuum system.}
	\end{minipage}%

\end{figure}

As the existing vacuum cell at Strathclyde \cite{kresic18} has a very limited trapping volume, a cell (see Fig.~\ref{Fig:cell_scheme})was designed to allow better optical access and wider trapping beams of 35~mm diameter which can be still delivered via 2-inch optics. The inner free aperture of the cell was chosen to be slightly larger, 50 mm. In order to prevent deformation of the cell under vacuum and a corresponding deformation of the optical phase fronts, 5~mm thick plates of fused silica are used. If the cavity is used at normal or close to normal incidence in order to prevent astigmatism affecting symmetry breaking, the trapping beams need to be injected at oblique angles. This lead to the design of a cell with rectangular surface area and outer dimensions of $160\times60\times60$~mm with a trapping volume of about 25~cm$^3$. We anticipate that this will enhance the number of atoms to be trapped by about a factor of 10 (to about $8-9\times 10^9$) and the on-resonance optical density by a bit more than a factor of 2 to 55-60.

For versatility, costs and complexity, the cavity will be placed outside the vacuum chamber. This enables in particular switching easily from a cavity to a single-mirror feedback configuration.  As the cavity will include four glass-air interfaces, without an anti-reflection coating the achievable finesse would be limited to less than 20 even for perfect mirrors. For a residual reflectivity of 0.1\%, the best achievable finesse is 780. Assuming a symmetric cavity with $R=0.994$, one obtains ${\cal T}=-\ln{R_{eff}}=0.01$ used in the discussion on the linear stability analysis and a finesse of about 300. Cavity length is dictated by cell size and will be about 75-100~mm. This results in a cavity linewidth of around 5-6~MHz comparable to the atomic linewidth and slightly larger than the stability of typical external cavity lasers. Reducing the finesse will reduce the requirements on laser stability and the linear stability analysis indicated that there is some flexibility to operate at lower finesse.

The configuration chosen for the cavity is a confocal one, for a number of reasons. It enables a high mode degeneracy important for multi-mode dynamics \cite{lippi93,kollar17} and the deviation from confocality constitutes an effective diffractive length. The symmetric confocal resonator is also quite stable and easy to align to minimize alignment and diffraction losses. However, odd and even modes are not degenerate. Hence it might be useful to consider to implement a quasi-planar cavity with a short diffractive length via intra-cavity telescopes. As the linear stability analysis indicates that the thresholds do not become prohibitive even for a finesse on the order of a only a few 10s, the additional losses might be acceptable.

\subsection{Assembly of vacuum cell}
As it is quite difficult to source glass cell with custom sizes and an anti-reflection coating on the inside as well as the outside, the decision was taken to assemble the cell in-house. High quality (10-5 scratch-dig and a specification of less than $\lambda/10$ transmitted wavefront distortion over the central 4 mm) fused silica windows of 5~mm thickness were obtained from Manx Precision Optics. The six pieces of glass were assembled using epoxy adhesive EPO-TEK 353ND. It was baked in an oven at 90-100$ ^\circ $C for one hour. The assembled cell is depicted in Fig.~\ref{Fig:cell_pic}. Afterwards it was connected to the rest of the vacuum system and pumped to high vacuum by a pumping rig system consistenting of a mechanical pump and a turbo pump.
The apparatus was baked for a week using heating tapes at 80-90$^\circ$C for the cell and at 95-115$^\circ$C for the rest of the system.
After finishing the procedure and sealing off the valve to the pump rig, the vacuum is sustained by a Vacion Plus 40 Starcell ion pump. The current running through the ion pump is stable and low, indicating that there are no outgassing elements within the vacuum. The monitor reading of the pump indicates a reading of a few times 0.1 $\mu$A implying a pressure around $p\approx 10^{-9}$ mbar according to the data sheet. This indicates good prospects to create a magneto-optical trap inside.

\section{Conclusion}
The linear stability analysis indicates that even for a cavity with a moderate finesse of order 300, threshold can be reached at easily accessible atom numbers, saturation parameters and high atomic detuning to minimize absorption losses. In the strongly detuned regime, the saturation threshold is decreasing inversely to on-resonance optical density. It is dominated by the optomechanical nonlinearity, not the electronic 2-level nonlinearity. The length scales of the emerging structures is controlled by cavity length and cavity detuning. In order to minimize extra-cavity pump intensity the latter one should not bee too high, but the numbers indicate that there is a substantial leeway to deal with additional losses at the glass-air interface due to a degradation of the AR-coating, imperfect alignment, intra-cavity lenses and higher cavity detunings.  The requirements are compatible with a flexible setup in which the cavity is situated outside the vacuum. A glass cell has been assembled allowing good optical access, which is suitable for cavity as well as single-mirror feedback experiments.

\ack{The equations used in the linear stability analysis in Sec.~\ref{sec:lsa} were derived by E. Tesio in his thesis \cite{tesio14t}. This work was performed in the framework of the European Training Network ColOpt, which is funded by the European Union (EU) Horizon 2020 programme under the Marie Skłodowska-Curie action, grant agreement 721465.

\section*{References}
\bibliographystyle{iopart-num}
\bibliography{sp_atom_crys_diff_deph_cavs.bib}

\end{document}